\documentclass[12pt,thmsa]{article}

\input tcilatex
\QQQ{Language}{
American English
}

\begin{document}

\begin{center}
Nature of Gravitation

Dr. A.V. Rykov

Chief of Seismometry lab. of IPE RAS, Moscow, Russia.
\end{center}

\textit{The photoeffect, (vacuum analogue of the photoelectric effect,) is
used to study the structure of the physical vacuum, the outcome of which is
the basis for an hypothesis on the nature of gravitation and inertia. The
source of gravitation is the vacuum which has a weak massless elementary
electrical dipole (+/-) charge. Inertia is the result of the elastic force
of the vacuum in opposition to the accelerated motion of material objects.
The vacuum is seen as the source of attraction for all bodies according to
the law of induction.}

The nature of gravitation remains one of the central problems of science and
the discovery of its true basis will introduce major changes to our
understanding of the physical laws. The following hypothesis is a departure
from commonly accepted physical theories. Newton presented the laws of
gravitation and inertia and shows acceleration as absolute in ambient space.
Einstein's General Theory represents gravitation as the curvature of space
near gravitating masses, and inertia is seen as equivalent to gravitation.
In consideration of the absolute or relative character of acceleration,
Einstein adopted Mach's theory, in which the property of inertia is seen as
the gravitational attraction of all masses in the Universe. This is despite
the paradox that an isolated rotating object should not experience
centrufugal forces. It is commonly acknowledged in physics, that the
curvature of space - time is sufficient, and gravitation is not required.
However this concept is not convincing even from a philosophical point of
view. The physics of the past century has continued the methodology of prior
centuries which is to search for answers to problems of HOW? and not WHY?,
considering the latter approach to be religious rather than physical. For
example, the Big Bang generates the whole substance of the Universe from a
mathematical point, presumably under no influence other than God. The
theories view the physical vacuum as playing an exclusive role in all
interactions except gravitation. Exchange forces are implemented in the
vacuum with the help of virtual particles: photons in electromagnetic
interaction, mesons in nuclear forces and gluons in nucleons. Gravitons as
exchange field quanta, have not received sufficient development in the
quantum theory of gravitation although a similar approach to the above is
indicated. The nature of gravitation is here presented as the vacuum
composed of massless charge dipoles, one component having a small charge
superiority over the other. In this manner, it is possible to represent a
primitive scheme of universal gravity and antigravity:

\begin{center}
( body1 +) ($-$ + $-$ + $-$ vacuum $-$ + $-$ + $-$) (+ body2 )
\end{center}

- The Coulomb attraction (gravity) in the presence of material bodies,

\begin{center}
($---$ vacuum $---$ )
\end{center}

- The Coulomb self-repulsion (antigravity) in the absence of material bodies
or bodies separated by large distances in space. An inequality zero of the
sum of charges is shown visually: [($-$) is numerically greater than (+).]
The retio of gravitation and repulsion in the universe forms the numerical
equality of $\Lambda $ -member in Einstein's theory [1, P. Davis, 1985].

At first we'll remove a blunder of physics presented by Coulomb's formula.
It lies in the fact that parameters of vacuum were put to the denominator of
formulas for electric and magnetic forces. We'l introduce inverse values:

$\eta =\frac 1\mu =1.0000000028\cdot 10^7[a^2kg^{-1}m^{-1}s^2]$ - is a
magnetic constant of vacuum equal to inverse value of magnetic permeability. 
$\xi =\frac 1\varepsilon =8.98755179\cdot 10^9[a^{-2}m^3kg\cdot s^{-4}]$ -
is a dielectric constant of vacuum equal to inverse value of dielectric
permittivity. Newton's and Coulomb's formulas get an identical view. Speed
of light gets more logical idea $c=\sqrt{\eta \xi }$ .

Experimental physics presents necessary data for the study of vacuum. We
mean the data on photoeffects in vacuum, on nuclei and nucleons [2,Karjakin
N.I. and others, 1964]. Let's remind the values of gamma-quanta energies: 1,
137, 1836, 3672 MeV ($2m_ec^2,137\cdot 2m_ec^2,1836\cdot 2m_ec^2,1836\cdot
4m_ec^2$). This series of energy gives a valuable information for the
physical ideas about the structure of vacuum and matter [3,Rykov A.V., 2001].

Gamma-quantum of $\nu $ frequency deforms the structure of cosmic vacuum.
Being within the size of $r_e$ between its elements, gamma-quantum creates a
deformation $\Delta r_e$ . The deformation energy will be $e_oE\Delta r_e$ $%
, $ where $e_o$ is a elementary charge, $E$ - is electrical intensity of the
structure. To avoid a well known experimental noise at real birth of
electron+positron pair by gamma-quantum we shall take the equation of the
energy in pure case:

\begin{center}
$h\nu =e_oE\Delta r_e$ (1),
\end{center}

where $h$ - is a Plank's constant. Deformation is function of time

\begin{center}
$\Delta r_e=\Delta [r_e\sin (2\pi \nu t)]=2\pi \nu r_e\Delta t\cos (2\pi \nu
t)$ (2).
\end{center}

Let's define the intensity of electrical field, where N is some coefficient
of proportionality:

\begin{center}
$E=N\xi \frac{e_o}{r_e^2}$ (3).
\end{center}

Let's put the obtained expressions, amplitude from (2) and intensity from
(3) to (1):

\begin{center}
$h=2\pi Ne_o^2\xi \frac 1{r_e/\Delta t}$ (4).
\end{center}

We can assume quite naturally that $r_e/\Delta t=c$ - is speed of light.
Let's find an unknown quantity:

\begin{center}
$N=\frac h{2\pi e_o^2r_q}=137.035990905=\alpha ^{-1}$ ! (5),
\end{center}

where $r_q=\sqrt{\xi /\eta }$ . We have got a well known formula of Plank's
constant:

\begin{center}
$h=2\pi e_o^2\alpha ^{-1}\sqrt{\xi /\eta }=6.6260755(40)\cdot 10^{-34}$ (6).
\end{center}

On this stage we should clear a situation with a choice of numerical values
for $h$ or $\alpha ^{-1}$ . All next values are calculated on the base of $h$%
. But the $\alpha ^{-1}$ is in reality more fundamental then $h$, because
the last one is derivative from $e_o^{},\alpha ^{-1},\xi $ $,\eta $ - vacuum
parameters. The choice made here is based upon this quite new study of
vacuum.

Gamma--quantum of energy $w\geq 1$ MeV interacting with vacuum changes a
''virtual'' electron-positron pair to the real ones. The energy equation of
this change is:

\begin{center}
$w=h\nu _{rb}=\xi \frac{e_o^2}{r_e}$ (7),$^{}$
\end{center}

where $r_e$ - distance between charges (+) and (-) of vacuum structure, $\nu
_{rb}=2.4892126289\cdot 10^{20}$ Hz - ''red border'' for frequency of
gamma-quantum . The last exact value is determined below. Let's find $r_e$ :

\begin{center}
$r_e=\frac{\xi \alpha }{2\pi r_q\nu _{rb}}=\frac{c\alpha }{2\pi \nu _{rb}}%
=1.398763188\cdot 10^{-15}m$ (8).
\end{center}

We have from (2) $\Delta r_e=2\pi \nu _{rb}r_e\Delta t=\frac{2\pi \nu
_{rb}r_e}cr_e=\alpha \cdot r_e$ under assumption $r_e/\Delta t=c$. In other
words, it is the limit of the vacuum deformation above what a rupture of
structure ties occurred:

\begin{center}
$\Delta r_e=\alpha \cdot r_e=1.020726874\cdot 10^{-17}m$ (9).
\end{center}

The exact value for $\nu _{rb}=\frac c{2\pi r_e\alpha ^{-1}}=2.48921263\cdot
10^{20}Hz$ . Deformation of structure lower than the given value has
electroelastic character. Let's find the coefficient of elasticity $b$ from
a next equation:

\begin{center}
$f=b\Delta r_{rb}=\xi \frac{e_o^2}{r_e^2}$ $,$ $b=1.155219829\cdot
10^{19}[kg\cdot s^{-2}]$ (10).
\end{center}

Dipoles can be polarized and the polarizetion will be next:

\begin{center}
$\sigma _{\Delta r}=\alpha ^{-2}\frac{e_o}{4\pi r_e^4}(\Delta r)^2=S(\Delta
r)^2$ , where

$S=\alpha ^{-2}\frac{e_o}{4\pi r_e^4}=6.254509137\cdot 10^{43}[Q\cdot m^{-4}$
(11).
\end{center}

Another useful parameter of vacuum will be :

\begin{center}
$E_\sigma =\sqrt{\gamma \xi }=0.77440463$ $[a^{-1}m^3s^{-3}]$ (12) .
\end{center}

The names for this parameters are not yet known.

To that stage we get the main parameters of the vacuum structure. Massless
vacuum structure follows the fact that energy required for creation pair of
electron+positron definites by energy equation $w=2m_oc^2+$ $2m_oc^2/137.036$
, where $2m_oc^2$ went on birth of two particle masses and $2m_oc^2/137.036$
went to break the dipole tie.

Dielectric vacuum media has a tied charges. The moving charge generates a
Maxwell's displacement current $j$. This current generates magnetic strength 
$\overline{dH}=\frac 1c\overline{j}\stackrel{}{}$ where $\overline{j}=\frac
1{4\pi }\frac{d\overline{E}}{dt}$. The $\overline{H}$ is necessary magnetic
component to the $\overline{E}$ for the Electromagnetic wave (light). The
vacuum structure is natural media for light excitation and propagation in
space. Thus, the connected charges - dipoles - are re-translators of an
electromagnetic wave. Light reaching the observer is not the initial
phenomenon of a photon emitted at source, but must be viewed as a
multiply-relayed signal.

It is natural to assume that the longitudinal polarization of the dipoles of
space involves gravitational phenomena. Gravitation is explained by the
electrostatic ''field'', which is transmitted in vacuo as a longitudinal
signal. The longitudinal motion of the polarized front between connected
charges is not accompanied by the appearance of a parallel magnetic field
moving in one direction, and of identical sign.The magnetic strength should
in this case surround the displacement current of moving charges similar to
a current in a conductor. As an electrostatics or gravitation act as central
and frequently spherical forces, the total magnetic strength of displacement
currents appears equal to zero for gravitating objects or those charged by
static electricity. The outcome is minimal damping. This infers an extremely
large and almost instantaneous speed of propagation of longitudinal waves in
the vacuum. The universe appears to be an interconnected system in which any
part ''feels'' in full unity with the whole. It is only in this way that it
is capable of existence and development. In essence, cosmology cannot manage
without ''instantaneous'' gravitational transfer.

The laws of Newton and Coulomb can be united next way.

$f=G\frac{m^2}{R^2}=\xi \frac{q^2}{R^2}$ and $\rho =\sqrt{\frac G\xi }%
=8.6164135164\cdot 10^{-11}[Q\cdot kg^{-1}]$- the electrical charge of one
kg of any mass. The same value may be presented by a micro parameters - $%
\rho =e_o\sqrt{\frac{2\pi G}{ch\alpha }}=8.6164135\cdot 10^{-11}$ $.$%
Gravitational constant is defined by parameters of vacuum $G=\xi \frac{e_o^2%
}{m_x}=6.67259049725\cdot 10^{-11}$[kg$^{-1}$m$^3$s$^{-2}$] where $m_x=m_{Pl}%
\sqrt{\alpha }=1.8594480544\cdot 10^{-9}$kg, $m_{Pl}$ - Plank mass. It is
indirect evidence of electrical nature of gravitation.

It is necessary to state that it is impossible to formally transfer accepted
physical concepts regarding material substance on the structured vacuum as
here indicated. Strength $E=\xi \frac q{R^2}$ and potential $U=\xi \frac qR$%
. For example, the calculation of acceleration of gravity for the earth in
terms of electrical forces gives $g=\sqrt{G\xi }\frac{\rho \,M}{R^2}$ . For
instance, The Earth has $g=9,82$ m/c\symbol{94}2 and electrical strength $%
E=1.1402\cdot 10^{10}$ V/m in \textit{vacuum}. This is nonsense from the
usual point of view. However, it is not surprising that the electrical
strength of an electron is $1.8367*10^{20}$ V/m and proton $6.399*10^{26}$
V/m. This is the medium in which '' the microparticles exist '', and of
which the macro bodies consist. Distances between the constituents of atoms
on 3-4 oder exceed the indicated distance. The vacuum penetrates everywhere,
whether it is a dielectric or a conductor. Therefore it must be realized
that customary concepts of shielding or electrical voltage are here
completely unsuitable. It is impossible for example, to arrange a conductor
between gravitating bodies to shield the operation of gravity. It is
impossible to arrange electrodes in space to remove and use the electrical
voltage of the vacuum. The carriers of electricity in a substance and in the
vacuum are completely different. The interaction of bodies with the vacuum
is implemented at the level of electrons and nucleons of substances.
Gravitation also begins at the same level, finally being integrated in
macroscopic masses.

We state that the force of elastic electrical deformation will be defined as

\begin{center}
$f=b\Delta r_{rb}=\xi \frac{e_o^2}{r_e^2}$ and $b=1.155065\cdot 10^{19}$[kg/s%
\symbol{94}2]. (13)
\end{center}

Where $b$ is the factor of electrical elasticity. Charge polarization -

\begin{center}
$\sigma =Q/4\pi R^2$ [Q/m\symbol{94}2](14).
\end{center}

Using formula (11), (14) and $g=G\frac M{R^2}$ for the acceleration of
gravity we have:

\begin{center}
$g=4\pi \sqrt{G\xi }S(\Delta r_g)^2$ m/s\symbol{94}2. (15)
\end{center}

The longitudinal deformation of the vacuum dipoles by a gravitating object
determines the acceleration of gravity and alternately, the acceleration of
gravity determines the deformation of the vacuum structure. We calculate the
maximum acceleration on (15)and (9):

\begin{center}
$g_{\max }=6.3409\cdot 10^{10}$ m/s\symbol{94}2. (16)
\end{center}

The force of electroelastic deformation from (9) will be defined by the
maximum acceleration of an unknown mass $m_x$:

\begin{center}
$b\Delta r_{rb}=g_{\max }m_x$ . (17)
\end{center}

The unknown mass is determined from the equation(17)

$m_x=\sqrt{\alpha }m_{Pl}=1.859459\cdot 10^{-9}$ kg, where $m_{Pl}$ -
Planck's mass !

This gives $Q=\rho m_x=1.602177\cdot 10^{-19}$ - the value of the charge of
an electron (!), inadvertently identifying a surprising connection of values 
$\rho ,\alpha ,m_x,m_{Pl}$, and that indirectly supports the gravitational
theory.

Mass provides the ability to determine the gear of gravitation through the
availability of a gravitational charge. We now calculate the number of pairs
of electrons and positrons forming vacuum dipoles in this mass: $%
n=m_x/m_o=2.0412553^{21}$ pieces. From this, the value of the charge is
determined $\Delta e_o=e_o/n=7.848981^{-41}$ Q, where the charge of an
electron exceeds the positron charge by: $7.848981^{-41}$ Q. For instance,
the excess of negative charges over positive by a factor of 21 is the basis
for gravitation. It corresponds to a minimum gravitational charge of the
mass of an electron or positron, i.e. $q_g=\rho m_o=$ $7.848981^{-41}$ Q. We
get there another very vivid coincidence and thus additional prove of
validity of description of nature gravity.

\begin{center}
Summary.
\end{center}

For centuries the nature of gravity is being unknown. The gravity was and is
up to now the most mysterious forth of the Nature. It is difficult to make
references to many-many publisched attempts to solve the problem of
gravity.It seems to auther that he find a realystic and physicaly based
theory of non-geometric gravitation. Above there are remarkable and
unexpected coincidences that serves as source of auther hope that this
article has sense. Any new theory should expect new knowledge about The
Nature.

1)The velocity of gravity is expected almost to be infinitiv. Experiment on
solar tide data in comparison with local sun time should give the estimated
velocity of gravity.

2)There are possibilites to control the deformation of vacuum structure by
electrical and magnetic forces, by gamma-quanta radiation etc. Thus the
gravity and inertia may be controled as the russian scientists Roschin V.V.
and Godin S.M. shows [4,Roschin,\ Godin, 2000]. There is a tale about
wonderful discs after John R.R.Searl which was taken by the mentioned
scientists for construction of their device.    

\begin{center}
Literature
\end{center}

1. Davis P. Superforth // Publisher ''Mir'', M., 1989, 277 p.

2. Karjakin N.I. et al. Physics Handbook // Publisher ''High School'', M.,
1964, 574 p.

3.Rykov A.V. Principles of natural physics // UIPE RAS, M.:2001, 58 p.(in
russian).

4.Roschin V.V., Godin S.M. Experimental research of physical effects in
dynamic magnetic system // The letters to MTP, St.Pb, 2000, v.26, is.24,
73-81 p.(in russian).

\end{document}